\documentclass{ecai}
\usepackage{times}
\usepackage{graphicx}
\usepackage{latexsym}

\ecaisubmission   % inserts page numbers. Use only for submission of paper.
\begin{document}

\title{A Turing Test for Crowds}

\author{Jamie Webster \and Martyn Amos\institute{Department of Computer and Information Sciences, Northumbria University, Newcastle upon Tyne, NE8 1ST, United Kingdom. Email: \{jamie.webster, martyn.amos\}@northumbria.ac.uk} }

\maketitle
\bibliographystyle{ecai}

\begin{abstract}
The realism and believability of crowd simulations underpins computational studies of human collective behaviour, with implications for urban design, policing, security and many other areas. Realism concerns the closeness of the fit between a simulation and observed data, and believability concerns the human perception of plausibility. In this paper, we ask two questions, via a so-called ``Turing Test" for crowds: (1) Can human observers {\it distinguish} between real and simulated crowds, and (2) Can human observers {\it identify} real crowds versus simulated crowds? In a study with student volunteers ($n$=384), we find convincing evidence that non-specialist individuals are able to reliably {\it distinguish} between real and simulated crowds. A rather more surprising result is that such individuals are overwhelmingly unable to {\it identify} real crowds. That is, they can tell real from simulated crowds, but are unable to say which is which. Our main conclusion is that (to the lay-person, at least) {\it realistic crowds are not believable} (and vice versa).
\end{abstract}

\section{Introduction}

The formal study of human crowds dates back to before the French Revolution \cite{drury2011contextualising}, but understanding collective behaviour is more urgent than ever before, as populations migrate to high-density urban centres, protests become more organized (and perhaps more common), and increasing numbers of individuals pass through large-scale transportation hubs \cite{drury2015crowds}. A number of computational techniques exist to study the dynamics of crowd behaviour, but the most commonly-used is {\it simulation} \cite{thalmann2013}.

Crowd simulations (generally, but not exclusively, using an agent-based approach) are now employed in many different domains, from events planning and management ~\cite{crociani2016multi}, to urban design ~\cite{feng2016crowd}, and incident response and analysis ~\cite{harding2011mutual,pretorius2015large}. By studying flows of people {\it en masse}, and their interactions with the environment and with one another, researchers aim to better understand human collective social behaviour, design more effective and enjoyable public spaces, and improve levels of safety, security and well-being \cite{halatsch2009value}.

In this paper, we consider two related properties of crowd simulations; (1) {\it realism}, and (2) {\it believability}. The first property concerns how well a simulation's output matches the expected or observed behaviour of a {\it real crowd} in the same scenario, under the same conditions. The second property centres on how {\it convincing} a simulation is to a human observer, and how closely it matches their {\it expectations} of how a crowd will behave. The two properties are closely inter-linked, and ``believability" is often used as a synonym for ``realistic". However, as we will see, the two concepts require close examination and careful handling.

The rest of the paper is organized as follows: in Section ~\ref{sec:bg} we briefly review related work on crowd simulation and collective behaviour; in Section ~\ref{sec:tt} we present our ``Turing Test" for crowd behaviour, and in Section ~\ref{sec:results} we give the results of experimental trials. We conclude with a brief discussion of the implications of our findings.

\section{Background}
\label{sec:bg}

The study of human crowd dynamics \cite{adrian2019glossary} is motivated by the desire to understand and predict the behaviour of individuals {\it en masse}, and encompasses a diverse range of crowd types, from large, mainly static crowds at sporting events or concerts \cite{wagner2014agent}, to transitory and flowing crowds, such as those found in train stations at rush hour \cite{zhang2008modeling}, or at religious events such as Hajj \cite{curtis2011virtual}. As urban centres grow in size (the United Nations predicts that, by 2050, 68\% of the global population will live in cities \cite{united2018}), we will need to understand and mitigate the impact of crowds on infrastructure, safety, security, and quality of life \cite{aschwanden2011empiric}. 

Early attempts to understand crowd behaviour were rooted in the physical sciences, using metaphors and mathematical tools from {\it fluid dynamics} \cite{henderson1974fluid}, and modelled crowds at the macroscopic level (i.e., without considering individuals) \cite{hughes2003flow}. Subsequent work used an {\it entity-based} approach, which treated crowds as individual ``particles" \cite{bouvier1997crowd,treuille2006continuum}, along with the {\it agent-based} methodology, in which individuals are treated as semi-autonomous actors \cite{pan2007multi}.

As crowd simulations have become used more frequently, attention has become focused on issues of {\it realism} and {\it believability}. Here, we define the realism of a simulation in terms of its {\it validity} \cite{johansson2007specification,klupfel2007simulation,pettre2009experiment,seer2014validating}; how closely does the output of the model match data obtained in the real world? It is straightforward to obtain statistical properties of simulation outputs and compare these to the properties of real-world crowds, and that is the approach we take in this paper.

The issue of {\it believability} is subtly different, and concerns the human perception of whether or not a crowd's behaviour is {\it plausible}. As computer-generated imagery (CGI) becomes increasingly common in large-scale cinematic productions, it is being used to replace human actors in large-scale crowd scenes, for reasons of cost and/or feasibility (e.g., the 2001 film {\it The Lord of the Rings: The Fellowship of the Ring} featured a prologue battle with 100,000 computer-generated fighters). Many modern video games also feature large numbers of simulated individuals. In these cases, the main concern is to ``fool" the observer into thinking that they are watching a real crowd, without necessarily producing patterns that are behaviourily valid \cite{durupinar2009ocean,peters2009modeling}.

\begin{figure}
\centerline{\includegraphics[height=2.5in]{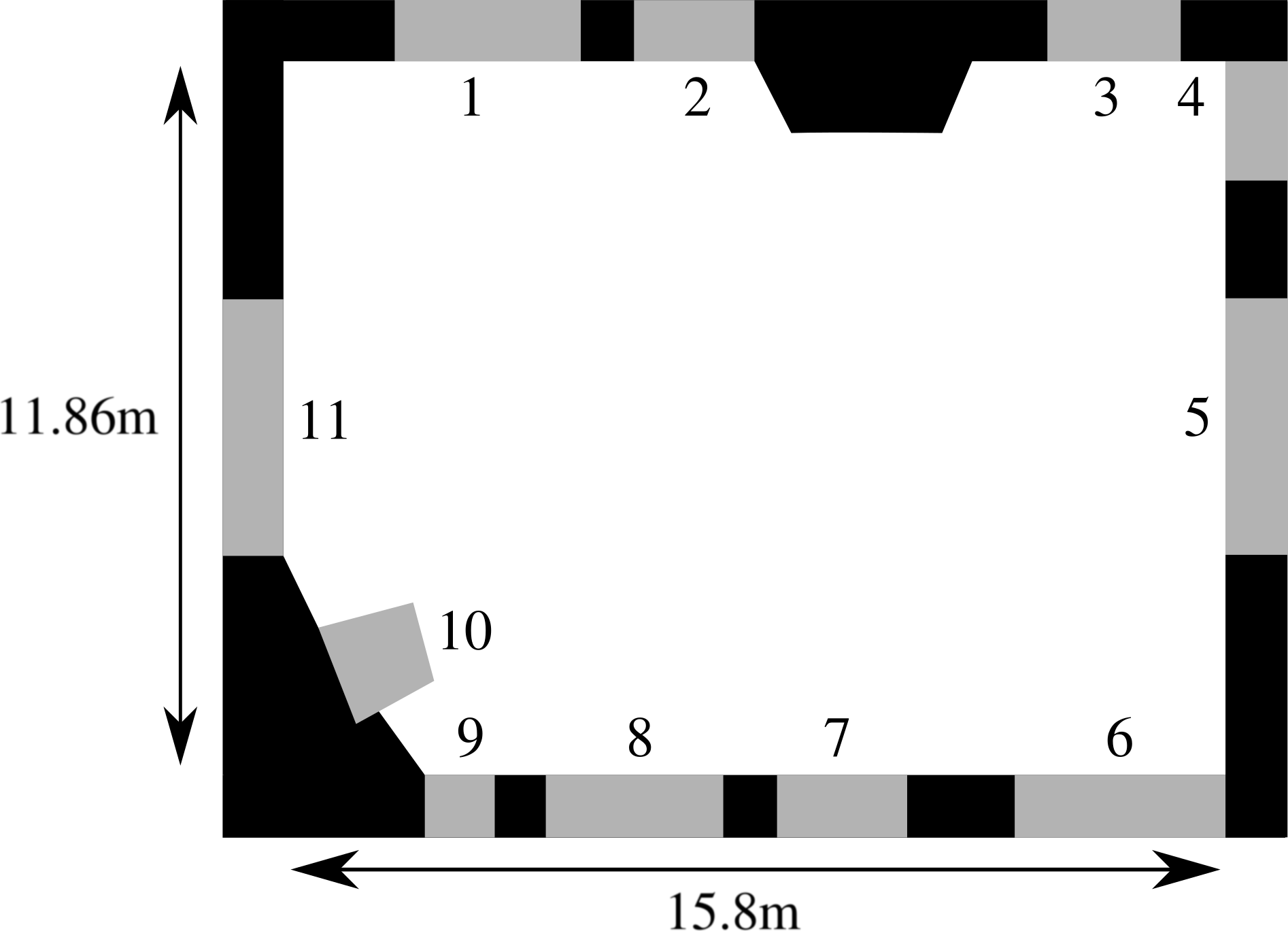}}
\caption{Diagram of Edinburgh Informatics Forum (ingress and egress points numbered).} \label{ForumDia}
\end{figure}

We seek, therefore, to study whether or not human observers may be persuaded that a simulated crowd  is actually a ``real" crowd. This may be thought of as a limited form of the famous ``Turing Test", named after Alan Turing, and described in his landmark paper on artificial intelligence \cite{machinery1950computing}. Turing proposed that if a human observer was unable to distinguish between another person and a machine designed to produce human-like responses in a conversational setting, then the machine would be deemed to have ``passed" the test. This type of test has been proposed for biological modelling \cite{harel2005turing} and artificial life \cite{cronin2006imitation} as a way of capturing and interrogating life-like properties of artificial systems, and assessing the completeness and validity of a model. Recently, a Turing test for collective motion in fish was described \cite{kn:Romenskyy2015}, and we base our approach on this. Our overall aim is to explore how a Turing-like test may be used to examine assumptions and preconceptions about the behaviour of human crowds, and to establish the features of real crowds that must be emulated by a simulation in order for it to be valid and/or ``pass" the test. In the next Section, we describe our methodology for doing so.

\section{Experimental method}
\label{sec:tt}

Our methodology is based on that of \cite{kn:Romenskyy2015}, but with {\it in-person} (as opposed to online) participants. We showed all participants six pairs of videos, in which one randomly-selected video showed the movement of a real crowd, and the other showed the movement of a simulated crowd. Both real and simulated crowd movements were displayed using the same rendering method, and participants were asked to specify on a form which of each pair they thought was the {\it real} crowd.

\subsection{Dataset}

\begin{figure}[t]
\centerline{\includegraphics[height=2in]{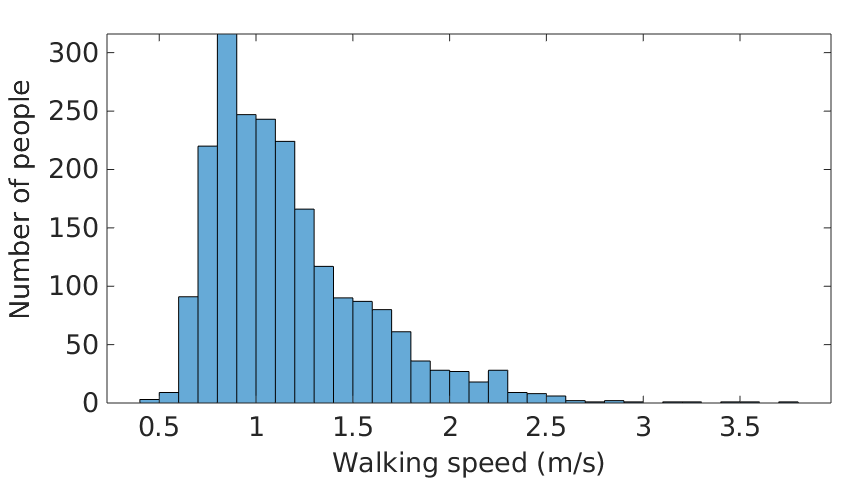}}
\caption{Distribution of walking speeds for pedestrians observed in Edinburgh Informatics Forum.} \label{speeds}
\end{figure}

We use data on real pedestrians from the University of Edinburgh School of Informatics ~\cite{kn:Majecka2009}. This public dataset, captured in 2010, contains over 299,000 individual trajectories corresponding to the movement of individuals through the School Forum, and is one of the largest open datasets of its type. A diagram of the Forum space is shown in Figure~\ref{ForumDia}. The Forum is rectangular in shape (measuring approximately 15.8 $\times$ 11.86 metres), has eleven ingress/egress points, and is generally clear of obstructions. Images were captured (9 per second) by a camera suspended 23m above the Forum floor, from which individual trajectories (``tracks") were extracted and made available (extraction was performed by the author of ~\cite{kn:Majecka2009}). This dataset has been used in a large number of studies of pedestrian movement/tracking, including \cite{fernando2018tracking,lovreglio2017towards,wang2016globally}, Importantly, none of the individuals whose trajectories were captured were actively participating in movement studies; the trajectories, therefore, are as close to ``natural" as possible (i.e., they have ``behavioural ecological validity" \cite{lovreglio2017towards}).

In what follows, we use the term ``clip" to specifically refer to a time-limited sequence of trajectory data (whether taken from the Edinburgh dataset or from the output of a simulation), as opposed to a movie visualisation. We wrote a utility to search the Edinburgh dataset and extract clips of a specific duration containing a specific number of individuals. This allowed us to ensure that the ``real" and ``simulated" crowds contained the same number of individuals for any single comparison.

\subsection{Model calibration}

In order to calibrate our simulation (and, later, to perform statistical analysis), we selected 20 clips at random from the Edinburgh dataset (each of 60s duration), and calculated the average walking speed of pedestrians observed traversing the Forum. The distribution of speeds is shown in Figure ~\ref{speeds}, with a mean value of 1.17m/s. When we simulated these scenarios (see next Section), the mean speed of agents was higher (1.63m/s), due to the fact that simulated agents were rarely impeded, did not encounter bottlenecks, and were free to accelerate up to their maximum speed. However, as we will see from our results, this did not affect the perception of the simulated crowds.

For the comparison experiments, we randomly selected six clips taken from the Edinburgh dataset (the number of clips is the same as in ~\cite{kn:Romenskyy2015}; each was of 60s duration, and the number of individuals in a clip ranged from 104 to 194 (with an average of 139). For each clip, we extracted the route choice distribution and the entry time distribution for all individuals.  This allowed us to initialise our simulations with the same distributions, ensuring that the runs closely matched the macroscopic properties of the real-world observations (while leaving room for the microscopic differences in which we are interested). We also calculated the average velocity of individuals in each clip, and used this to scale the clip's length (by modifying the video playback speed) to account for variability in camera capture rate, thus normalizing the velocity of individuals relative to expected walking speed~\cite{kn:Bohannon1997}.

\newpage
\subsection{Simulations}

In order to produce the simulations to accompany each Edinburgh clip, we simulated pedestrian movement using the Vadere package ~\cite{kn:Kleinmeier2019}. This package is open-source, which means that (unlike commercial software) its movement models are open to inspection, and it also allows for easy exporting of simulating pedestrian trajectories (which is important when we consider that we must use the same renderer for both real and simulated videos). 

A crucial component of the simulation is the {\it crowd motion model}. This defines the rules of interaction between individuals (e.g., avoidance), and between individuals and their environment (e.g., repulsion from walls and physical obstacles), as well as route choice behaviour and differential walking speed. Many different crowd motion models exist \cite{duives2013state}, but perhaps the most commonly-used type is based on {\it social forces}. Inspired by the fluid flow paradigm of Henderson \cite{henderson1974fluid} and others, Helbing and Molnar's social force model (SFM) ~\cite{kn:Helbing1995} is a microscopic, continuous model which uses ``attractive" and ``repulsive" force fields between individuals (and between individuals and their environment) to guide movement. The SFM provides the base movement model for a number of pedestrian simulation packages, including FDS+EVAC \cite{korhonen2010fds}, PedSim \cite{kn:Gloor16}, SimWalk \cite{kimura2003pedestrian} and MassMotion \cite{massmotion}, and it has been used extensively in movement research. Additionally, the SFM has been validated using real-world data \cite{johansson2007specification,seer2014validating}, and the comprehensive review of \cite{duives2013state} recommends its use in pedestrian movement studies. For all simulations, we used the pre-supplied Vadere template for Helbing and Molnar's SFM, with default attributes and parameters (listed in Table~\ref{Params}). 

\begin{table}
\begin{center}
{\caption{Vadere simulation model parameters.}\label{Params}}
\begin{tabular}{lccccccc}
\hline
\rule{0pt}{12pt}
Parameter&Value
\\
\hline
\\[-6pt]
\quad ODE Solver & Dormand-Prince method\\
\quad Pedestrian body potential & 2.72\\
\quad Pedestrian recognition distance & 0.3\\
\quad Obstacle body potential & 20.1\\
\quad Obstacle repulsion strength & 0.25\\
\quad Pedestrian radius (m) & 0.2\\
\quad Pedestrian speed distribution mean (m/s) & 1.4\\
\quad Pedestrian minimum speed (m/s) & 0.4\\
\quad Pedestrian maximum speed (m/s) & 3.2\\
\quad Pedestrian acceleration (m/s) & 2.0\\
\quad Pedestrian search radius (m) & 2.0\\
\\
\hline
\\[-6pt]
\end{tabular}
\end{center}
\end{table}

We added small amounts of noise to the simulated trajectories in order to replicate noise in the real crowd data. As the Edinburgh individuals were detected by an overhead camera running at 9fps, occasional faulty detections caused very short-term errors in the extracted trajectories. Once rendered, this cause individuals to appear to rapidly ``flick" between two headings. As we had no reliable way to quantify the (by inspection, small) amount of noise in the trajectories, we adjusted this by eye until the apparent noise in the simulated data matched the noise level observed in the real data. At any time-step, a simulated agent had a 15\% chance of temporarily ``flicking" their heading by a randomly selected value up to 45 degrees (without changing their trajectory). Importantly, as we will see from the results, the addition of this noise had no effect on how the simulated crowds were perceived. 

A second artefact of inaccurate detections was that some trajectories had missing sections for several time steps; once rendered, these individuals would temporarily disappear from the frame and then reappear. To fix this, we automatically detected such situations and interpolated coordinates for the missing time-steps when parsing the Edinburgh dataset. We also increased the number of frames per second of both sets of trajectories (real and simulated), from 9 to 72, by interpolating coordinates. This enabled smooth video playback for the purpose of comparisons.

\subsection{Comparisons}

\begin{figure}
\centerline{\includegraphics[height=2.3in]{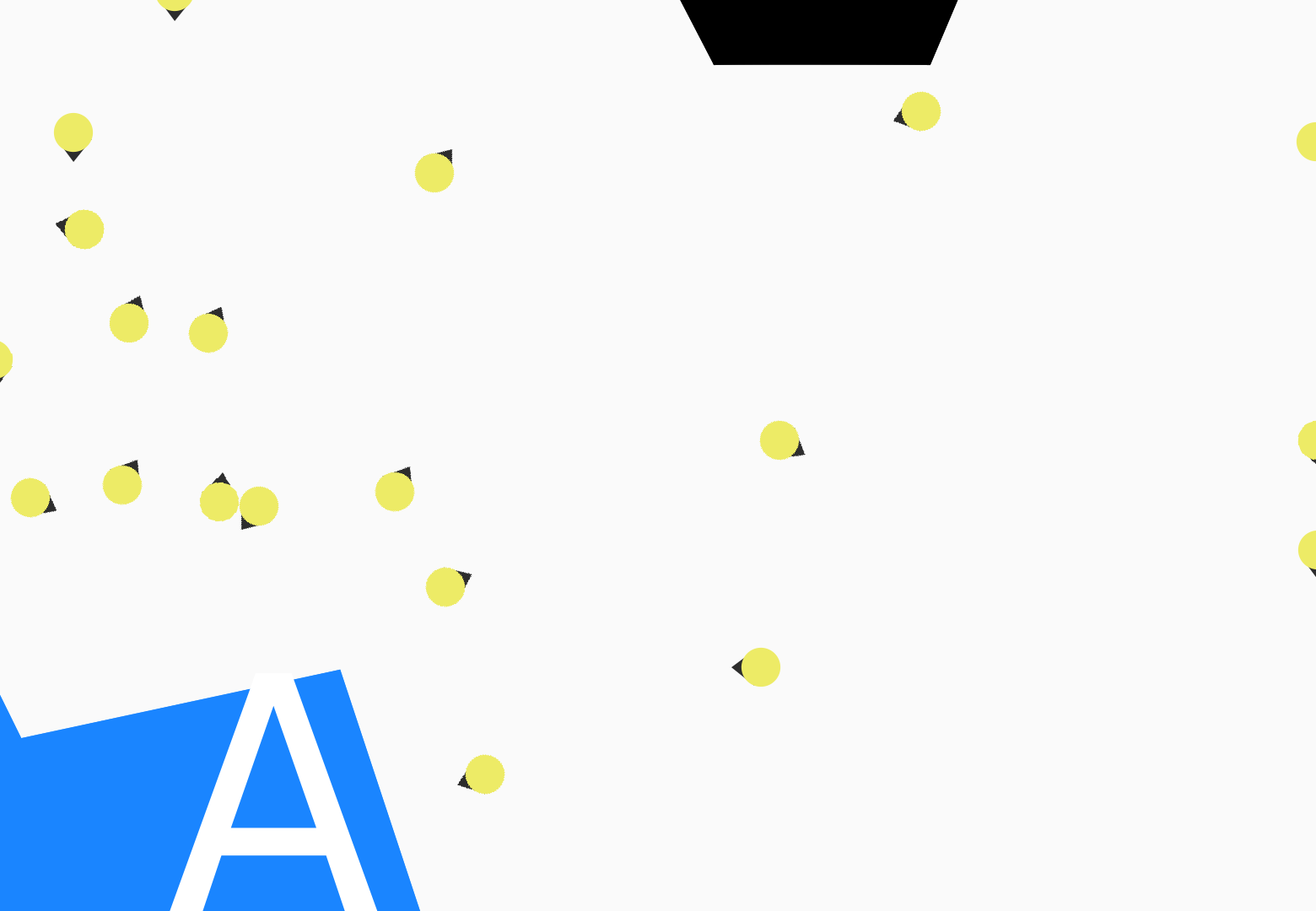}}
\caption{Single frame render of an example crowd.} \label{render}
\end{figure}

The trajectories of both the simulated and real individuals in each pair of clips were then rendered in a uniform fashion, using a tool coded in Java. This allowed us to produce ``top down" visualisations of both real and simulated clips that were uniform in appearance, with individuals represented as filled circles, and headings depicted by an arrow (see Figure~\ref{render}).

For each pair, the real and simulated videos were randomly assigned to position A (left) or B (right), and these were combined side-by-side into a single video. Individual videos did not ``loop", and were made up of the first 30 seconds of the real and simulated crowd clips in each pair. The total duration of the video, showing a total of six comparisons, was 3m 18s (including a 3s pause between each pair). The video is available at http://drives.media/google857, and the real crowds are A, A, B, A, B and B.

\subsection{Statistical properties}

The first question we asked was whether or not participants could distinguish between real and simulated crowds, even when their statistical properties were very similar. That is, are there features of real (or simulated) crowds that somehow act as an discriminator, even when there is no significant statistical difference between the two?

We used two metrics (as in ~\cite{kn:Romenskyy2015}); {\it polarization} and {\it nearest neighbour distance} (NND). Polarization measures the level of ``order" in a crowd, in terms of the heading alignment of members. Polarization is zero when the crowd is completely disordered (everyone is pointing in a different direction), and has a maximum value of 1 when all members of the crowd have the same heading:

\begin{equation}
    \varphi = \left\langle \frac{1}{N}  \left| \sum_{N}^{i=1} \exp(\iota\theta_{i}) \right|  \right\rangle,
\end{equation}
 
where $N$ is the number of individuals in the frame, $\iota$ is the imaginary unit, and $\theta_{i}$ is the heading of each individual.

Nearest-neighbour distance (NND) measures the level of ``clustering in a crowd.
The average NND for a single ``frame" (derived from either the real dataset or the simulation) is calculated from the sum of nearest-neighbour distances of all $N$ individuals:

\begin{equation}
    \nu = \frac{1}{N} \sum_{N}^{i=1} d_{i},
\end{equation}

where d$_{i}$ is the nearest neighbor distance between point $i$ and the closest individual in the frame, as calculated by the standard distance formula,

\begin{equation}
    d_{i} = \sqrt{(x_{2} - x_{1})^2 + (y_{2} - y_{1})^2}
\end{equation}

\begin{figure}[t]
\centerline{\includegraphics[height=1.9in]{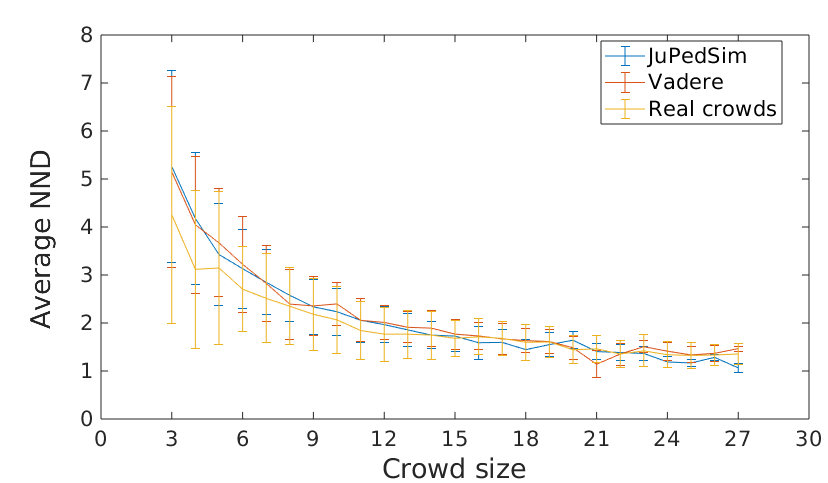}}
\centerline{\includegraphics[height=1.9in]{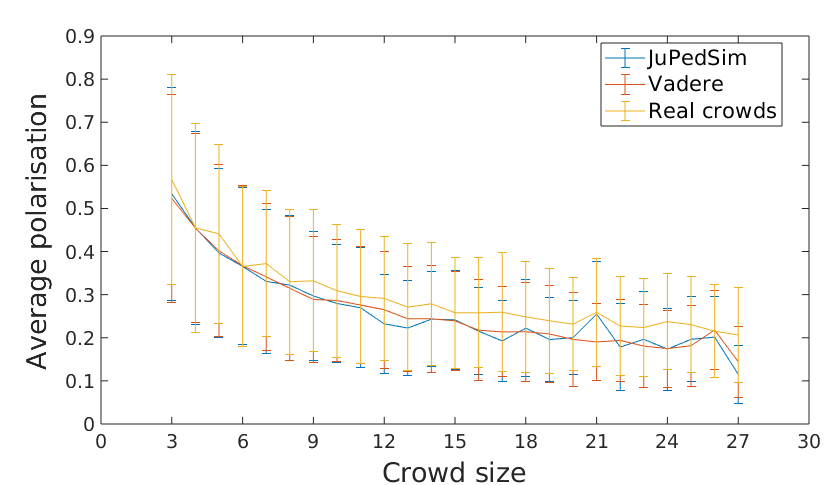}}
\caption{Crowd simulations/real crowd comparisons: Nearest Neighbour Distance (NND) (top) and polarization (bottom) as a function of crowd size. The outputs of both simulations have statistical properties that are close to those of the real crowds.} \label{Represent}
\end{figure}

In order to confirm that we did not introduce implementation-specific bias by choosing a specific software platform, we compared the outputs of Vadere and JuPedSim ~\cite{kn:Wagoum15}, an alternative open-source simulation package. We used each package to simulate the 20 real crowd clips mentioned in the previous Section, and calculated average NND and polarization over 20 runs for each. The same statistics were then calculated for the real clips (Figure ~\ref{Represent}).

These results confirmed that crowd simulations in Vadere and JuPedSim display similar properties in terms of both NND and polarization, so we selected Vadere as a representative example of crowd simulation packages in general. 

Importantly, the statistical properties of the simulations also matched the general properties of the {\it real} crowds, which confirmed that they are essentially indistinguishable in those terms. In Figure ~\ref{Represent} we notice a slight difference between the real and simulated crowds in terms of polarization; the real crowds are generally slightly more closely-aligned than the simulated crowds, but this difference is on the order of 2\%, and we do not believe that this is significant enough to introduce any perceptible difference.

\section{Experiments}
\label{sec:results}

We recruited 384 undergraduate students from Northumbria University, distributed over 9 groups taken from a mixture of computer science and engineering courses. Of the participants who supplied their details, the gender distribution was 78.83\% male, 18.66\% female, 2.5\% non-binary/other, and the average age was 20.7 (we exclude one outlier age value of 71, corresponding to a student's reader). All trials took place at the beginning of a class, for which prior permission was obtained from the tutor. Students were informed about the nature of the experiment, and told that they were under no obligation to participate. Answer sheets were distributed, which consisted of a simple numbered list of tickboxes (for each line, the choice was A or B). Participants were asked to optionally specify their age and gender. At the end of the trial, participants were also asked to provide some optional narrative notes on any distinguishing features they noticed that allowed them to tell the real crowd from the simulated crowd. Each trial (from initial set-up to collection of answer sheets) took around ten minutes. The experimental protocol was approved by the Northumbria University Faculty of Engineering and Environment Ethics Board, application number 16433.

 \subsection{Results}

\begin{figure}
\centerline{\includegraphics[height=1.9in]{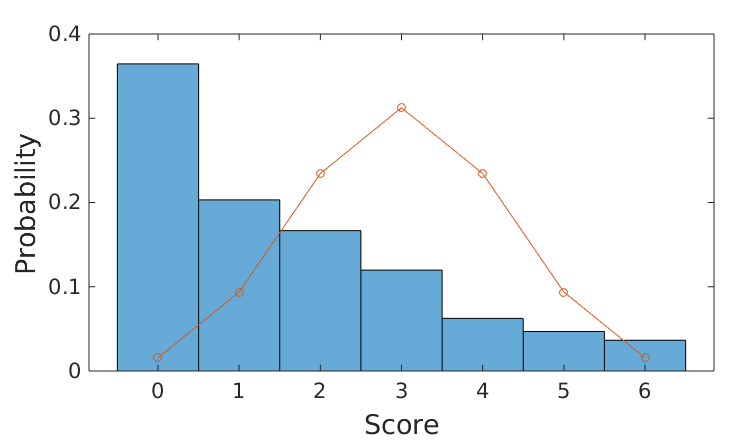}}
\caption{Distribution of participant scores (the line represents the expected binomial distribution).} \label{overall}
\end{figure}

We define ``score" in terms of correct identification of the real crowd; so a score of zero means that a participant failed to identify any of the real crowds, and a score of 6 means that the participant identified the real crowd on every occasion. The average score for participants, across all comparisons, was 1.6. That is, participants performed less well than if they had guessed at random. The overall distribution of scores is shown in Figure ~\ref{overall}, overlaid with the expected binomial distribution (as each comparison is a binary choice, we show this to illustrate the expected distribution of scores if selections were made at random)\footnote{Full results are available at https://doi.org/10.6084/m9.figshare.10308407}. 

If the real and simulated crowds were genuinely indistinguishable (that is, the best strategy would be no better than random guessing), then we would expect roughly 3\% of participants (around 12 people) to either guess none correctly, or to guess all six correctly. What we actually found was that over 40\% of participants (154 individuals) obtained a score of either zero or six. That is, those individuals were able to correctly {\it partition} all six pairs of videos into two sets. This answers, in the affirmative, the first research question: can individuals {\it distinguish} between real and simulated crowds, even when they have very similar statistical properties?

A highly striking result is that the most common score, by far, was zero. That is, a significant proportion of participants (36.46\%) failed to identify a single real crowd. Only 3.65\% of participants obtained a perfect score of 6. The important implication of this is that participants were reliably able to {\it partition} videos along the lines of ``real"/``artificial", but most of them were unable to say which was which. This is a much stronger version of the result obtained in \cite{kn:Romenskyy2015}, where participants were able to tell real fish from simulated fish, but were not necessarily able to identify the real fish. 

We conclude, therefore, that the second research question (can individuals {\it identify} real crowds versus simulated crowds, even when they have very similar statistical properties?) must be answered in the negative for this population. In the next Section, we analyse the narrative text supplied by participants, in order to explore the underlying rationale for their decisions.

\begin{figure}
\centerline{\includegraphics[height=1.9in]{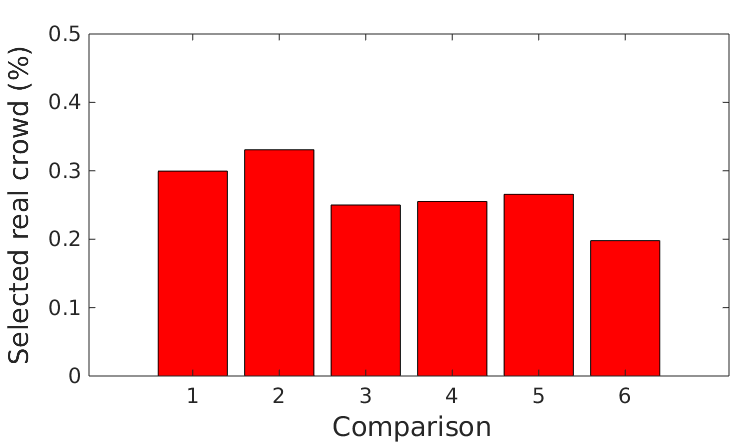}}
\caption{Success rate across individual pairwise comparisons.} \label{crosscomp}
\end{figure}

We now briefly explore secondary features of our findings. The results for each comparison are shown in Figure~\ref{crosscomp}, which we present in terms of the proportion of participants who correctly selected the real crowd. These results show that pair 6 presented the strongest challenge to participants, and pair 2 was considered the least challenging. Overall, no clear trend emerged in terms of differential challenge across comparisons.

\begin{figure}
\centerline{\includegraphics[height=1.9in]{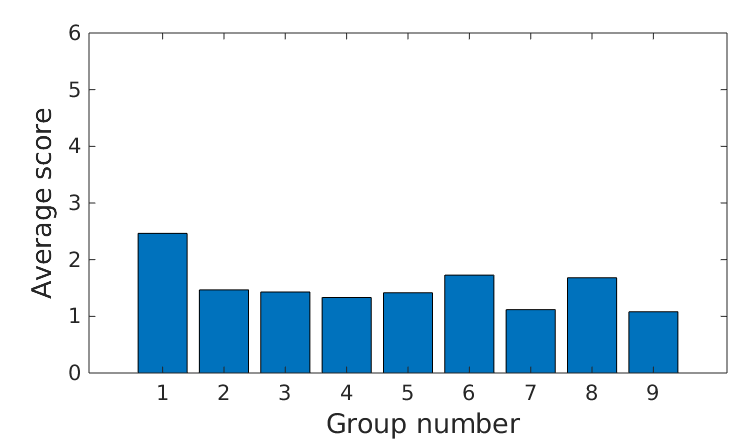}}
\caption{Distribution of scores across participant groups.} \label{crossgroup}
\end{figure}

In terms of variation across groups (Figure ~\ref{crossgroup}), Group 1 (second year Engineering Mathematics students) obtained the most correct identifications, with an average score of 2.46. Group 9 (first year computer science students) had the fewest correct identifications, with an average score of 1.08.

\subsection{Narrative findings}

In this Section, we analyse the free text supplied by participants. We focus, in particular, on the large number of participants who scored zero, as they consistently misidentified the real crowd. We highlight themes and specific comments that may shed light on the assumptions and preconceptions held by these individuals, that led them to consistently ``flip" the real and simulated crowds in their perception.

The first theme that emerged concerned rapid or ``random" {\it changes of movement} in the real crowd, which many participants incorrectly attributed to the simulated (``fake") crowd. Versions of this included ``Fake changed direction too quickly", ``Fast change suggests fake", ``Generated crowd had too much random movement", ``Real seemed to change direction gently". Although the average speed of the simulated agents was higher than that of the real people, participants singled out rapid movement in the {\it Edinburgh} videos as indicative of artificiality (when, in fact, the real people moved more slowly). Overall, 72 participants mentioned a variant of this type of observation. The underlying assumption here is that real people move smoothly, at a uniform speed, and do not tend to deviate much from their chosen path.

A second common theme concerned {\it avoidance}; many participants incorrectly assumed that real people would avoid close contact with one another, whereas the simulated individuals would ``overlap" or collide. Representative quotes included ``Simulated people collided, real crowds avoided each other", and ``People overlapping". In reality, the opposite is true, as the real dataset contains multiple instances of individuals coming into close proximity. Moreover, the social forces model {\it explicitly} tries to keep individuals {\it apart} unless close proximity is unavoidable, so the behaviour (distance keeping) that participants attributed to real people was actually an in-built feature of the simulation. This theme was mentioned by 22 participants.

Perhaps the most profound observations concerned {\it perceived intentionality} and {\it group-level behaviour}; many participants believed that ``On the whole, people have relatively smooth and intentional paths" (this was actually a feature of the simulation), ``Real crowds don't really stand around" (stationary groups were only present in the real dataset), and ``The real ones knew where they were going" (this was actually a function of the simulation's path choice algorithm). Variations on this theme were mentioned by 7 participants. The interesting thing here is that participants (incorrectly) ascribed clear human intentionality and purpose to the {\it simulated} agents (``Real crowds move more purposefully"), and failed to acknowledge it in the actual humans that were observed.

Overall, we found that participants believed that individuals in crowds are orderly, purposeful,
respectful of personal space, and consistent and uniform in their speed and direction. In fact, all of these characteristics were features of the simulation. Participants also failed to recognize features of real crowds such as rapid changes in speed or direction, close proximity of individuals, and stationary groups/individuals, all of which were discounted by participants as being ``glitchy" or ``unrealistic". Our findings would, therefore, appear to contradict \cite{lerner2009fitting}, in which the authors state that ``However, people do more than just walk. They talk to one another, they look
around, they scratch their heads or perform various other actions... The absence of these
mundane actions diminishes the credibility of the simulated crowd." Such arguments about ``diminished credibility" would seem to have little empirical basis in fact, and even when attempts are made to integrate ``realistic" behaviours into crowd simulations, no rigorous test is performed to assess whether or not they have had any impact on perceived plausibility in human observers.

\section{Conclusions}

In this paper, we presented a Turing test for crowds that allowed us to investigate issues of realism and believability in crowd simulations by comparing them with visualisations representing data obtained from real pedestrians. We performed trials with nearly 400 university students, and found that, while the students were generally able to discriminate between ``real" and ``artificial" crowds, they were unable to say which was which. On the surface, this was a rather surprising result, but it serves to emphasise that a simulation model that is {\it realistic} (that is, accurately reflects reality) might be useful for planning an evacuation scenario, where fidelity is paramount, but it might be less useful where ``believability" is more important (in a video game, for example). Conversely, this serves as a warning that software development platforms aimed primarily at entertainment titles should perhaps not be relied upon for safety-critical or potentially expensive infrastructure studies without careful modification.

We acknowledge several potential limitations of our study; the use of students as test subjects is the subject of ongoing debate \cite{peterson2001use}, and the computer science background of many of the students (and the gender imbalance) may have biased our results. It may be the case that our students have become conditioned to make certain assumptions about how crowds behave from playing games that use a relatively unrealistic crowd model. However, this is merely speculation on our part. Nonetheless, an important future development of this work will be to rerun the trials using experts in crowd dynamics, to find out whether they are better placed to identify the real crowd. This is entirely consistent with Harel's expectation of how a biological modelling Turing test might work; ``...our interrogators can't simply be any humans of average intelligence. Both they and the .... people responsible for `running' the real organism and providing its responses to probes, would have to be experts on the subject matter of the model, appropriately knowledgeable about its coverage and levels of detail." \cite{harel2005turing}

If (as we might expect) the experts are able to reliably identify the real crowd, then this immediately suggests a mechanism for ascertaining the minimal set of crowd features that are necessary to ``pass" the test. If, for example we identified that ``group-level movement" was a ``flag" for the experts, we might include such a behaviour in the simulation and rerun the trial with a second group of experts. If the success rate falls, and the experts are less able to tell the difference between real and simulated crowds, then we might conclude that group-level behaviour constitutes an important feature that should be included in simulation packages. This would represent a formalised methodology for implementing a number of recommendations that have been recently made by a number of crowd scientists, who call for the integration into software of a wider range of psychological and inter-personal processes \cite{Lemercier2016,Seitz2017,Templeton2015}. These recommendations reflect a pressing need to revisit physics-based models of crowd behaviour which, though they may generate macroscopic behaviour that is reasonably realistic, fail to capture the inherent ``messiness" and unpredictability of real human crowds.

\ack We thank Jeremy Ellman for advice on statistical analysis.
\newpage
\bibliography{ecai.bib}{}

\begin{thebibliography}{10}

\bibitem{adrian2019glossary}
J.~Adrian, M.~Amos, M.~Baratchi, et~al., `A glossary for research on human
  crowd dynamics', {\em Collective Dynamics}, {\bf 4}(A19),  1--13, (2019).

\bibitem{aschwanden2011empiric}
G.D.P.A. Aschwanden, S.~Haegler, F.~Bosch{\'e}, L.~Van~Gool, and G.~Schmitt,
  `Empiric design evaluation in urban planning', {\em Automation in
  Construction}, {\bf 20}(3),  299--310, (2011).

\bibitem{kn:Bohannon1997}
R.W. Bohannon, `Comfortable and maximum walking speed of adults aged 20—79
  years: reference values and determinants', {\em Age and Ageing}, {\bf 26}(1),
   15--19, (1997).

\bibitem{bouvier1997crowd}
E.~Bouvier, E.~Cohen, and L.~Najman, `From crowd simulation to airbag
  deployment: particle systems, a new paradigm of simulation', {\em Journal of
  Electronic Imaging}, {\bf 6}(1),  94--108, (1997).

\bibitem{crociani2016multi}
L.~Crociani, G.~L{\"a}mmel, and G.~Vizzari, `Multi-scale simulation for crowd
  management: a case study in an urban scenario', in {\em International
  Conference on Autonomous Agents and Multiagent Systems (AAMAS)}, pp.
  147--162. Springer, (2016).

\bibitem{cronin2006imitation}
L.~Cronin, N.~Krasnogor, B.G. Davis, C.~Alexander, N.~Robertson, J.H.G.
  Steinke, S.L.M. Schroeder, A.N. Khlobystov, G.~Cooper, P.M. Gardner, et~al.,
  `The imitation game—a computational chemical approach to recognizing life',
  {\em Nature Biotechnology}, {\bf 24}(10),  1203, (2006).

\bibitem{curtis2011virtual}
S.~Curtis, S.J. Guy, B.~Zafar, and D.~Manocha, `Virtual {T}awaf: A case study
  in simulating the behavior of dense, heterogeneous crowds', in {\em 2011 IEEE
  International Conference on Computer Vision Workshops (ICCV Workshops)}, pp.
  128--135. IEEE, (2011).

\bibitem{drury2011contextualising}
J.~Drury and C.~Stott, `Contextualising the crowd in contemporary social
  science', {\em Contemporary Social Science}, {\bf 6}(3),  275--288, (2011).

\bibitem{drury2015crowds}
J.~Drury and C.~Stott, {\em Crowds in the 21st Century: Perspectives from
  Contemporary Social Science}, Routledge, 2015.

\bibitem{duives2013state}
D.C. Duives, W.~Daamen, and S.P. Hoogendoorn, `State-of-the-art crowd motion
  simulation models', {\em Transportation Research Part C: Emerging
  Technologies}, {\bf 37},  193--209, (2013).

\bibitem{durupinar2009ocean}
F.~Durupinar, N.~Pelechano, J.~Allbeck, U.~Gudukbay, and N.I. Badler, `How the
  ocean personality model affects the perception of crowds', {\em IEEE Computer
  Graphics and Applications}, {\bf 31}(3),  22--31, (2009).

\bibitem{feng2016crowd}
T.~Feng, L-F. Yu, S-K Yeung, K.~Yin, and K.~Zhou, `Crowd-driven mid-scale
  layout design.', {\em ACM Transactions on Graphics}, {\bf 35}(4),  132--1,
  (2016).

\bibitem{fernando2018tracking}
T.~Fernando, S.~Denman, S.~Sridharan, and C.~Fookes, `Tracking by prediction: A
  deep generative model for multi-person localisation and tracking', in {\em
  2018 IEEE Winter Conference on Applications of Computer Vision (WACV)}, pp.
  1122--1132. IEEE, (2018).

\bibitem{kn:Gloor16}
C.~Gloor.
\newblock {PedSim}: Pedestrian crowd simulation, 2016.
\newblock http://pedsim.silmaril.org.

\bibitem{halatsch2009value}
J.~Halatsch, A.~Kunze, and G.~Schmitt, `{V}alue {L}ab: A collaborative
  environment for the planning of {F}uture {C}ities', in {\em Proceedings of
  eCAADe}, volume~27, (2009).

\bibitem{harding2011mutual}
P.~Harding, S.~Gwynne, and M.~Amos, `Mutual information for the detection of
  crush', {\em PloS ONE}, {\bf 6}(12),  e28747, (2011).

\bibitem{harel2005turing}
D.~Harel, `A {T}uring-like test for biological modeling', {\em Nature
  Biotechnology}, {\bf 23}(4),  495, (2005).

\bibitem{kn:Helbing1995}
D.~Helbing and P.~Molnar, `Social force model for pedestrian dynamics', {\em
  Physical Review E}, {\bf 51}(5),  4282, (1995).

\bibitem{henderson1974fluid}
L.F. Henderson, `On the fluid mechanics of human crowd motion', {\em
  Transportation research}, {\bf 8}(6),  509--515, (1974).

\bibitem{kn:Romenskyy2015}
J.E. Herbert-Read, M.~Romenskyy, and D.J. Sumpter, `{A Turing test for
  collective motion}', {\em Biology Letters}, {\bf 11},  20150674, (2015).

\bibitem{hughes2003flow}
R.L. Hughes, `The flow of human crowds', {\em Annual Review of Fluid
  Mechanics}, {\bf 35}(1),  169--182, (2003).

\bibitem{johansson2007specification}
A.~Johansson, D.~Helbing, and P.K. Shukla, `Specification of the social force
  pedestrian model by evolutionary adjustment to video tracking data', {\em
  Advances in Complex Systems}, {\bf 10}(supp02),  271--288, (2007).

\bibitem{kimura2003pedestrian}
T.~Kimura, H.~Sekine, T.~Sano, N.~Takeichi, Y.~Yoshida, and H.~Watanabe,
  `Pedestrian simulation system {SimWalk}', in {\em Summaries of Technical
  Papers of Annual Meeting Architectural Institute of Japan, E-1}, pp.
  915--916, (2003).

\bibitem{kn:Kleinmeier2019}
B.~Kleinmeier, B.~Z{\"o}nnchen, M.~G{\"o}del, and G.~K{\"o}ster, `Vadere: An
  open-source simulation framework to promote interdisciplinary understanding',
  {\em arXiv preprint arXiv:1907.09520}, (2019).

\bibitem{klupfel2007simulation}
H.~Kl{\"u}pfel, `The simulation of crowd dynamics at very large events -
  calibration, empirical data, and validation', in {\em Pedestrian and
  Evacuation Dynamics (PED) 2005},  285--296, Springer, (2007).

\bibitem{korhonen2010fds}
T.~Korhonen, S.~Hostikka, S.~Heli{\"o}vaara, and H.~Ehtamo, `{FDS}+{E}vac: an
  agent based fire evacuation model', in {\em Pedestrian and Evacuation
  Dynamics (PED) 2008},  109--120, Springer, (2010).

\bibitem{Lemercier2016}
S.~Lemercier and J.M. Auberlet, `{Towards more behaviors in crowd simulation}',
  {\em Computer Animation And Virtual Worlds}, (2016).

\bibitem{lerner2009fitting}
A.~Lerner, E.~Fitusi, Y.~Chrysanthou, and D.~Cohen-Or, `Fitting behaviors to
  pedestrian simulations', in {\em Proceedings of the 2009 ACM
  SIGGRAPH/Eurographics Symposium on Computer Animation}, pp. 199--208. ACM,
  (2009).

\bibitem{lovreglio2017towards}
R.~Lovreglio, C.~Dias, X.~Song, and L.~Ballerini, `Towards microscopic
  calibration of pedestrian simulation models using open trajectory datasets:
  the case study of the {E}dinburgh {I}nformatics {F}orum', in {\em Conference
  on Traffic and Granular Flow, Washington DC, USA}, (2017).

\bibitem{kn:Majecka2009}
B.~Majecka, {\em Statistical models of pedestrian behaviour in the forum},
  Master's thesis, School of Informatics, University of Edinburgh, 2009.

\bibitem{united2018}
United Nations.
\newblock 2018 {R}evision of {W}orld {U}rbanization {P}rospects, 2018.

\bibitem{massmotion}
Oasys.
\newblock Mass motion product page, 2019.
\newblock
  https://www.oasys-software.com/products/pedestrian-simulation/massmotion/.

\bibitem{pan2007multi}
X.~Pan, C.S. Han, K.~Dauber, and K.H. Law, `A multi-agent based framework for
  the simulation of human and social behaviors during emergency evacuations',
  {\em AI \& Society}, {\bf 22}(2),  113--132, (2007).

\bibitem{peters2009modeling}
C.~Peters and C.~Ennis, `Modeling groups of plausible virtual pedestrians',
  {\em IEEE Computer Graphics and Applications}, {\bf 29}(4),  54--63, (2009).

\bibitem{peterson2001use}
R.A. Peterson, `On the use of college students in social science research:
  Insights from a second-order meta-analysis', {\em Journal of Consumer
  Research}, {\bf 28}(3),  450--461, (2001).

\bibitem{pettre2009experiment}
J.~Pettr{\'e}, J.~Ond{\v{r}}ej, A-H. Olivier, A.~Cretual, and S.~Donikian,
  `Experiment-based modeling, simulation and validation of interactions between
  virtual walkers', in {\em Proceedings of the 2009 ACM SIGGRAPH/Eurographics
  Symposium on Computer Animation}, pp. 189--198. ACM, (2009).

\bibitem{pretorius2015large}
M.~Pretorius, S.~Gwynne, and E.R. Galea, `Large crowd modelling: an analysis of
  the {D}uisburg {L}ove {P}arade disaster', {\em Fire and Materials}, {\bf
  39}(4),  301--322, (2015).

\bibitem{seer2014validating}
S.~Seer, C.~Rudloff, T.~Matyus, and N.~Br{\"a}ndle, `Validating social force
  based models with comprehensive real world motion data', {\em Transportation
  Research Procedia}, {\bf 2},  724--732, (2014).

\bibitem{Seitz2017}
M.J. Seitz, A.~Templeton, J.~Drury, G.~K{\"{o}}ster, and A.~Philippides,
  `{Parsimony versus reductionism: How can crowd psychology be introduced into
  computer simulation?}', {\em Review of General Psychology}, {\bf 21}(1),
  95--102, (2017).

\bibitem{Templeton2015}
A.~Templeton, J.~Drury, and A.~Philippides, `{From mindless masses to small
  groups : conceptualizing collective behavior in crowd modeling}', {\em Review
  of General Psychology}, {\bf 19}(3),  215--229, (2015).

\bibitem{thalmann2013}
S.~Thalmann and S.R. Musse, {\em Crowd Simulation}, Springer, 2013.

\bibitem{treuille2006continuum}
A.~Treuille, S.~Cooper, and Z.~Popovi{\'c}, `Continuum crowds', in {\em ACM
  Transactions on Graphics (TOG)}, volume~25, pp. 1160--1168. ACM, (2006).

\bibitem{machinery1950computing}
A.M. Turing, `Computing machinery and intelligence', {\em Mind}, {\bf 59}(236),
   433, (1950).

\bibitem{wagner2014agent}
N.~Wagner and V.~Agrawal, `An agent-based simulation system for concert venue
  crowd evacuation modeling in the presence of a fire disaster', {\em Expert
  Systems with Applications}, {\bf 41}(6),  2807--2815, (2014).

\bibitem{kn:Wagoum15}
A.K. Wagoum, M.~Chraibi, J.~Zhang, and G.~Lammel, `{JuPedSim}: an open
  framework for simulating and analyzing the dynamics of pedestrians', in {\em
  3rd Conference of Transportation Research Group of India}, (2015).

\bibitem{wang2016globally}
H.~Wang and C.~O’Sullivan, `Globally continuous and non-{M}arkovian crowd
  activity analysis from videos', in {\em European Conference on Computer
  Vision}, pp. 527--544. Springer, (2016).

\bibitem{zhang2008modeling}
Q.~Zhang, B.~Han, and D.~Li, `Modeling and simulation of passenger alighting
  and boarding movement in {B}eijing metro stations', {\em Transportation
  Research Part C: Emerging Technologies}, {\bf 16}(5),  635--649, (2008).

\end{thebibliography}
\end{document}